\documentclass[amscd,amsmath,amssymb,verbatim, aps, prd, twocolumn,showpacs]{revtex4}

\usepackage{graphicx}
\usepackage[dvips]{hyperref}
\usepackage{textcomp}
\usepackage[OT1,T1]{fontenc}

\begin{document}

\title{Could the Pioneer anomaly have a gravitational origin?}
\author{Kjell Tangen}
\email{kjell.tangen@dnv.com}
\affiliation{DNV, 1322 H{\o}vik, Norway}\label{I1}
\date{\today}
\begin{abstract}
If the Pioneer anomaly has a gravitational origin, it would, according
to the equivalence principle, distort the motions of the planets
in the Solar System. Since no anomalous motion of the planets has
been detected, it is generally believed that the Pioneer anomaly
can not originate from a gravitational source in the Solar System.
However, this conclusion becomes less obvious when considering models
that either imply modifications to gravity at long range or gravitational
sources localized to the outer Solar System, given the uncertainty
in the orbital parameters of the outer planets. Following the general
assumption that the Pioneer spacecraft move geodesically in a spherically
symmetric spacetime metric, we derive the metric disturbance that
is needed in order to account for the Pioneer anomaly. We then analyze
the residual effects on the astronomical observables of the three
outer planets that would arise from this metric disturbance, given
an arbitrary metric theory of gravity.\ \ Providing a method for
comparing the computed residuals with actual residuals, our results
imply that the presence of a perturbation to the gravitational field
necessary to induce the Pioneer anomaly is in conflict with available
data for the planets Uranus and Pluto, but not for Neptune. We therefore
conclude that the motion of the Pioneer spacecraft must be non-geodesic.
Since our results are model independent within the class of metric
theories of gravity, they can be applied to rule out any model of
the Pioneer anomaly that implies that the Pioneer spacecraft move
geodesically in a perturbed spacetime metric, regardless of the
origin of this metric disturbance.
\end{abstract}
\maketitle

\section{Introduction}\label{XRef-Section-215221244}

\subsection{Brief summary of the Pioneer anomaly}

Precision tracking of the Pioneer 10 and 11 spacecraft over more
than 30 years has revealed an anomalous frequency drift of the Doppler
radar signals transmitted by the spacecraft\ \ \cite{Anderson:1998,Anderson:2001sg,Turyshev:2005vj,Turyshev:2005zm,Markwardt-2002}.\ \ After
careful analysis of potential sources of this frequency drift, it
has been concluded that it most likely is caused by an anomalous
acceleration of both spacecraft, directed towards the Sun. Moreover,
the anomalous acceleration seems to stay constant over long distances,
and it has been estimated to have very similar values for both spacecraft,
about $8\times {10}^{-10}m/s^{2}$over distances between 20 and 70
AU from the Sun \cite{Turyshev:2005vj}. The data presented by Anderson
et. al. \cite{Anderson:2001sg,Turyshev:2005zm} show that the anomalous
acceleration became noticable beyond the orbit of Saturn, i.e. around
10 AU. It quickly grew to a near constant value of around $8\times
{10}^{-10}m/s^{2}$ directed towards the Sun, and has remained more
or less constant beyond 20 AU. As the Pioneer anomaly remains unaccounted
for, the question has been raised whether the anomaly could originate
from unknown physics. Indeed, a number of proposals for possible
physical explanations have been made. These range from conventional
physics, such as anisotropic heat dissipation \cite{Scheffer:2003se,Olsen2006},
drag from interplanetary dust \cite{Nieto:2005zs} or gravitational
attraction from Kuiper-belt objects \cite{Nieto:2005bw}, to\ \ new
physics, such as scalar fields\cite{Bertolami-Paramos-2005}, brane-world
scenarios \cite{Bertolami-Paramos-2004}, dark matter, Modified Newtonian
Dynamics (MOND) \cite{Milgrom:1983ca,Bekenstein-Magueijo-2006},
modifications to the Newtonian gravitational potential \cite{Reynaud-Jaekel-2005},
modifications to General Relativity \cite{Jaekel-Reynaud-2005b},
and other alternative gravitational theories, see refs.\ \ \cite{Anderson:2001sg,Turyshev:2005vj}
and references therein.

A natural question to ask is whether the effect could have a gravitational
origin\footnote{It should be noted that, by {\itshape gravitational},
we here mean any physical interaction mediated indirectly through
the spacetime metric, regardless of the nature of its source or
how the metric couples to the source. A gravitational theory in
this context means any metric theory of gravity, and gravitational
sources would include not only matter or radiation, but also exotic
sources, like Dark Energy, scalar fields, as well as the 4-dimensional
effects of higher-dimensional models, such as the brane-world models.
}, either originating from an unknown gravitational source present
within the Solar System, or being a manifestation of long-range
modifications to gravity, i.e. arising from\ \ deviations in the
vacuum gravitational field from the field predicted by General Relativity,
where these deviations become observable only over longer distances.
If the latter were found to be true, it would imply the need for
a revision of established gravitational theory, a possibility that,
however remote it might be, by itself explains the attention the
Pioneer anomaly has received within the fundamental physics community.
If the Pioneer anomaly were indeed a gravitational effect, one\ \ would,
according to the equivalence principle, expect such a disturbance
in the gravitational field to also affect the motions of the planets.
The question of whether the Pioneer anomaly could be attributed
to a gravitational source in the Solar System was considered by
Anderson et. al. \cite{Anderson:2001sg}. The orbits of the inner
planets are determined with very high precision, and no anomalous
motion of these planets have been observed, so if a gravitational
source were to explain the Pioneer anomaly, it could not be present
in the inner Solar System.\ \ However, because the Pioneer effect
first became visible as the spacecraft entered the outer Solar System,
their analysis does not exclude the possibility of a gravitational
source localized to the outer regions of the Solar System, beyond
the orbit of Saturn. In this case, one\ \ would expect the disturbance
in the gravitational field created by such a source to only affect
the motions of the three outer planets, which orbit the Sun at distances
between 20 and 40 AU, but it would have no observable effect on
the motions of the inner planets. Since the orbits of the three
outer planets have been determined by optical methods alone\ \ \cite{Standish-1998,Pitjeva-2005a},
the uncertainty in their orbital parameters is much higher than
for the other planets. Recently, Page, Dixon and Wallin \cite{Page-Dixon-2006a},
Izzo and Rathke\ \ \cite{Izzo-Rathke-2005} and\ \ Iorio and Giudice
\cite{Iorio-2006a} examined the question whether the Pioneer anomaly,
given that it is a gravitational effect, could have an observable
effect on the orbits of the outer planets.\ \ Page et. al. concluded
that, if the Pioneer anomaly has gravitational origin, it would
not have an observable effect on the motions of the outer planets,
given the present accuracy of planetary data for these planets.\ \ Their
conclusion was based on the fact that\ \ only optical observations
have been made of Uranus, Neptune and Pluto \cite{Standish-1998,Pitjeva-2005a}\ \ and
that the a priori uncertainty of the optical measurements for the
three outer planets is substantial, resulting in rather large uncertainties
in their orbital parameters. Such uncertainties could possibly mask
the disturbances to the planetary orbits caused by an anomalous
gravitational field. However, no calculations were shown to substantiate
their conclusion. Izzo and Rathke, on the other hand, came to a
different conclusion. They concluded that the Pioneer anomaly could
not have gravitational origin, because it would be in conflict with
the ephemerides of Uranus and Neptune. They used parametric constraints
set on a parametrization of deviations from Newtonian gravity in
terms of an effective reduced Solar mass $\mu _{\odot }( r) $ \cite{Talmadge-1988}.
Applying the constraints to the planets Uranus and Neptune, they
concluded that the deviations to the Newtonian gravitational potential
implied by the Pioneer anomaly would lead to\ \ changes in the effective
reduced Solar mass that exceeded the limits set by Solar System
measurements by 1-2 orders of magnitude. Iorio and Giudice \cite{Iorio-2006a}
came to the same conclusion, providing an assessment of the\ \ perturbations
to the orbits of the outer planets based on the Gauss equations.
The Gauss equations estimate the effect of a perturbation in the
gravitational field in terms of the time rate of change it causes
on the osculating orbital elements. Iorio and Giudice found large
secular effects on the argument of the perihelion and mean anomaly
for all three outer planets.

The possibility that the Pioneer anomaly might indicate a deviation
in the vacuum gravitational potential from the one predicted by
General Relativity has also been studied. A possible Yukawa modification
to the Newtonian gravitational potential has been dismissed because
of constraints from planetary data \cite{Anderson:2001sg,Reynaud-Jaekel-2005}.
However, modified gravitational theories have been proposed that
are claimed to account for the Pioneer anomaly without violating
parametric constraints inferred from planetary data \cite{Jaekel-Reynaud-2005b}.

\subsection{What is the appropriate question?}\label{XRef-Subsection-41364936}

As we can see from this brief review of the subject (which by no
means is complete, for a more complete review see refs. \cite{Anderson:2001sg,Turyshev:2005vj}),
the number of models that have been proposed to explain the Pioneer
anomaly is already significant, and growing fast. Fitting a model
to parametric constraints is useful for dismissal and initial scrutiny
of models, but a closer examination of the models that pass the
parametric constraints is in place. When looking at the models aiming
to explain the Pioneer anomaly, they are different, but many of
them are based on the assumption that the motion of the Pioneer
spacecraft is geodesic. The question whether the Pioneer effect
could have gravitational origin, either being caused by an unknown
gravitational source localized in the outer regions of the Solar
System or arising from\ \ long-range deviations in the gravitational
field from the vacuum field predicted by General Relativity, can
be reformulated as a question of whether the Pioneer spacecraft
move geodesically. Within the class of metric theories of gravity,
this question is model independent. As will be seen, it can be satisfactorily
answered by giving quantitative, model-independent estimates derived
from the basic equations of motions, of the perturbations to the
orbits of the outer planets that would result from such an effect.
These estimates could then be compared with observations, thus providing
a more thorough test of models against observations. To settle this
question would be of significance for the substantial model building
effort targeting the Pioneer anomaly. Furthermore, if it were possible
to prove that the Pioneer spacecraft moved non-geodesically, it
would imply that we could rule out a large class of models, including
several of the models that have been proposed to explain the Pioneer
anomaly.

\subsection{And how can we answer it?}

The purpose of this paper is to derive an answer to the question
posed above; whether the Pioneers move geodesically. We will perform
a quantitative, model-independent assessment of the perturbations
to the motions of the outer planets that would arise from a disturbance
to the spacetime metric of the kind required to account for the
Pioneer anomaly. The rationale for analyzing the motions of the
outer planets is, in addition to the obvious reason that they orbit
the Sun at distances at which the Pioneer anomaly has been observed,
the relatively large uncertainty in their orbital parameters \cite{Standish-1998,Pitjeva-2005a}.
Therefore, the likelihood of our result will to a large extent depend
on this uncertainty.

The procedures that are currently applied for calculating the ephemerides
of the planets use available theory, taking into account known gravitational
sources, and carefully fitting the computed orbits to available
observational data \cite{Standish-1998,Pitjeva-2005a}. One would
therefore expect that any disturbance to the gravitational field
within the solar system would, if large enough, over time generate
residuals in the position measurements of the planets as they would
tend to drift from their expected positions. The question then is,
what residuals would arise if we were to explain the Pioneer anomaly
as the result of a perturbation to the gravitational field in the
outer solar system, and how do they compare with the observed residuals?

Whether or not this metric perturbation is induced by a real gravitational
source or is just a manifestation of a long-range deviation in the
vacuum gravitational field from the field predicted by General Relativity
is irrelevent for our study. In order to determine whether a perturbation
to the gravitational field has an observable effect on planetary
orbits, we will have to simulate the effect that the gravitational
perturbation would have on the planetary orbits (what we will refer
to as the {\itshape gross perturbations} or {\itshape gross effects}),
then compute the simulated ephemeris as the solution to the unperturbed
equations of motion that provides the best fit to the perturbed
solution, compute the simulated residual as the difference between
those two solutions and finally compare the simulated residuals
with the real residuals.

An assessment similar to our study was recently done by Iorio and
Giudice \cite{Iorio-2006a}, based on the Gauss equations for orbital
motion. The Gauss equations estimate rates of change of orbital
elements, an approach that makes the effects easy to understand.
On the other hand, the\ \ orbital elements are not directly observable,
and the estimation of observables from them is not straight forward.
In their work, they investigated the gross orbit perturbations caused
by a constant, anomalous acceleration of the Pioneer kind. They
found large secular effects on the argument of the perihelion and
mean anomaly for all three outer planets. As will be shown later,
such large secular effects will largely be canceled by the ephemeris,
because secular effects are lowest-order effects that to a large
extent will be matched by a best fit solution to the unperturbed
equations of motion.\ \

Our approach is to work from the basic equations of motion, thus
working with the real orbit of the body, not with approximate orbital
elements. Relating our results to measurements is only a matter
of making coordinate transformations between the reference frame
of the calculation and the reference frame of the observer. We apply
the hypothesis that the Pioneer spacecraft move geodesically through
space, i.e. that the only forces acting on them are gravitational
and that the anomalous acceleration is the effect of disturbances
to the spacetime metric. Furthermore, we assume a spherically symmetric
gravitational field that can be described within the framework of\ \ a
metric theory of gravitation. The spacecraft move approximately
in opposite directions and their trajectories have both minor inclinations
relative to the ecliptic plane \cite{Turyshev:2005vj}, so a spherically
symmetric gravitational field at large\ \ distances from the Sun
should be a reasonable assumption. Since we are assuming a metric
theory of gravitation, the equivalence principle will be valid within
the theory \cite{Will-2001,Bertolami-Paramos-Turyshev-2006}. At
the outset, we assume nothing about how the metric couples to gravitational
sources or of the nature of any gravitational source, so our results
will be model independent in the sense that they will only depend
on the assumptions of geodesic motion in a static spacetime metric
and spherical symmetry. We will work in the weak field limit, and
in this limit, a metric theory of gravitation can be parametrized
using the PPN formalism \cite{Will-2001,Bertolami-Paramos-Turyshev-2006}.
Thus, our results will be valid for any PPN-parametrizable gravitational
theory.

We are only interested in dominant effects and will work perturbatively
to lowest order in the non-relativistic weak field limit. To lowest
order, independent perturbations decouple, so we may disregard other
effects that perturb the planetary orbits, like disturbances from
other planets or satellites. This simplification allows us to treat
the problem analytically by doing perturbative expansions around
idealized\ \ analytical solutions to the planetary orbits.

\subsection{Article overview}

In section \ref{XRef-Section-4264724} we derive and solve the perturbative
equations of motion for Newtonian gravity. We make a quantitative
assessment of the gross perturbations in order to get a better understanding
of the real effects of the gravitational perturbation. In this section
we also account for the accuracy of the perturbative expansions
applied.

Then, in section \ref{XRef-Section-415172115}, we make the generalization
of this procedure to any metric theory of gravitation. First, we
derive the unique form that the perturbation to the gravitational
field must have in order to be able to account for the Pioneer anomaly.
From the equations of motion, it is possible to uniquely derive
the signature of the gravitational potential that would cause a
constant anomalous acceleration. Having uniquely constrained the
spacetime metric, we are able to derive the perturbative equations
for geodesic motion and find the general solution to them.

In section \ref{XRef-Section-4264518}, we cover in detail the analysis
method that will be employed in order to determine whether the perturbations
computed in sections \ref{XRef-Section-4264724} and\ \ \ \ \ref{XRef-Section-415172115}
are observable or not.

Finally, in section \ref{XRef-Section-426554}, we apply this analysis
method to the results of sections \ref{XRef-Section-4264724} and
\ref{XRef-Section-415172115}. Here, we first derive the simulated
ephemeris, which is the solution to the unperturbed equations of
motion that best fits the perturbed solution. Having obtained the
simulated ephemeris, we then compute the simulated residuals as
the difference between the observable values of the\ \ perturbed
orbit and the simulated ephemeris. Finally, we are able to conduct
the result analysis described in Section \ref{XRef-Section-4264518},
where we compare the simulated residuals with the real residuals.\ \

\section{Perturbations to planetary orbits -\ \ Newtonian gravity}\label{XRef-Section-4264724}

We will start our analysis with the simplest case, which is that
of a static, spherically symmetric, Newtonian gravitational field.

\subsection{Equations of motion}

For an object moving freely in a static, spherically symmetric gravitational
field, motion is planar. In order to describe the motion of the
object, let us therefore introduce in-plane coordinates $(r, \varphi
)$, where $r$ is radial distance from the central mass to the object,
and $\varphi $ measures the longitude of the object in the orbital
plane, relative to a fixed reference direction. In the case of planetary
motion, we will define $\varphi$ as the {\itshape argument of latitude},
which is positive in the direction of movement and measured relative
to the ascending node, the point where the body ascends through
the ecliptic plane.\ \ The equations of motion can easily be reduced
to the following form, expressed in terms of the resiprocal distance
$u\equiv 1/r$ \cite{Goldstein-1950}:
\begin{gather}
\frac{d\varphi ( t) }{dt}=\frac{\ell }{r^{2}}%
\label{XRef-Equation-313225824}
\\\frac{d^{2}u( \varphi ) }{d\varphi ^{2}}+u=-\frac{\mathcal{F}(
u) }{\ell ^{2}u^{2}},%
\label{XRef-Equation-313183311}
\end{gather}

where the constant $\ell $ is the specific angular momentum (angular
momentum per unit mass) and $\mathcal{F}( u) $ is the force per
unit mass. Any solution to these equations have two constants of
motion: The angular momentum $L$ and the total energy $E$, which
is the sum of kinetic and potential energy. For a Newtonian gravitational
field surrounding a central mass $M$, the force $\mathcal{F}( u)
$ has the familiar inverse square form: $\mathcal{F}_{N}=-\frac{G
M}{r^{2}}$, where $G$ is the gravitational constant. In this case,
eq. \ref{XRef-Equation-313183311} takes the form
\begin{equation}
\frac{d^{2}u( \varphi ) }{d\varphi ^{2}}+u=\frac{G M}{\ell ^{2}}%
\label{XRef-Equation-313195439}
\end{equation}

The bound solutions to eq. \ref{XRef-Equation-313195439}\ \ are
the familiar Keplerian orbits with elliptical form:
\begin{equation}
r_{0}( \varphi ) =\frac{1}{u_{0}( \varphi ) }\equiv \frac{a( 1-e^{2})
}{1+e \cos ( \varphi -\omega ) },%
\label{XRef-Equation-32618372}
\end{equation}

where $a$ is the semi-major axis of the orbit, $e$ is the eccentricity
and $\omega $ is a constant, the {\itshape argument of periapsis}
of the orbit. Notice that the angular difference $\varphi -\omega
$ is the {\itshape true anomaly} of the orbit{\itshape ,} denoted
$\nu$.

\subsection{Pioneer perturbation}\label{XRef-Subsection-32891218}

Let us perturb the Newtonian gravitational field with an anomalous,
constant radial accelerations $a_{p}=8.7\times {10}^{-10}m/s^{2}$
directed towards the Sun. The force $\mathcal{F}$ takes the form
\begin{equation}
\mathcal{F}=\mathcal{F}_{N}-a_{p}=-\frac{G M}{r^{2}}-a_{p}%
\label{XRef-Equation-31323220}
\end{equation}

Assume that $u_{0}$ is a reference solution to the unperturbed equations
of motion, of the form given by eq. \ref{XRef-Equation-32618372}.
Define $\omega _{0}$ to be the argument of periapsis of the reference
orbit. Then, let us define the perturbation $\delta u=u-u_{0}$,
where $u$ is a solution to the perturbed equation of motion, eq.
\ref{XRef-Equation-313183311},\ \ with the force $\mathcal{F}$ defined
by eq. \ref{XRef-Equation-31323220}.

From eq. \ref{XRef-Equation-313183311}, we obtain the following
equation of motion for the perturbation $\delta u$:
\begin{equation}
\frac{d^{2}\delta u}{{\mathrm{d\varphi }}^{2}}+\left( 1+\frac{2\kappa
}{{\left( 1+e \cos  \nu \right) }^{3}}\right) \delta u=\frac{\kappa
}{a( 1-e^{2}) {\left( 1+e \cos  \nu \right) }^{2}}%
\label{XRef-Equation-68193227}
\end{equation}

 where the constant $\kappa \equiv \frac{a_{p}}{a_{N} }{(1-e^{2})}^{2}$
is proportional to the ratio between the anomalous acceleration
$a_{p}$ and the mean Newtonian acceleration $a_{N}\equiv G M/ a^{2}$,
and we have introduced $\nu \equiv \varphi -\omega _{0}$ for notational
convenience. Evaluating $\kappa $ at heliocentric distances relevant
for this study yields $\kappa  \lesssim {10}^{-4}$. The presence
of $\kappa $ on the left hand side of eq. \ref{XRef-Equation-68193227}
gives rise to a small perihelion precession of the orbit. However,
the contribution from the $\kappa$-term is small compared with the
total perturbation. This can easily be seen by taking the circular
limit of eq. \ref{XRef-Equation-68193227}, in which case $\delta
u=\frac{\kappa }{a( 1+2\kappa ) }\approx \frac{\kappa }{a}$. For
eccentricities $e<1$, this will still be the case. Subsequently,
we will disregard the angular dependency of the $\kappa$-term on
the left hand side of eq. \ref{XRef-Equation-68193227} and use the
approximate equation \footnote{The validity of this approach has
been checked by comparing analytical and numerical solutions to
the perturbation equations}
\begin{equation}
\frac{d^{2}\delta u}{{\mathrm{d\varphi }}^{2}}+\left( 1+2\kappa
\right) \delta u=\frac{\kappa }{a( 1-e^{2}) {\left( 1+e \cos  \nu
\right) }^{2}},%
\label{XRef-Equation-32621731}
\end{equation}

Initial conditions for the perturbation are chosen by demanding
that the solutions $u$ and $u_{0}$ have the same angular momentum
$ \ell $ ($\delta \ell =0$). Furthermore, we will demand that the
perturbations $\delta u$ and $\delta$$\varphi$ vanish identically
in the limit $\kappa \rightarrow 0$, i.e. the perturbed orbit approaches
the reference orbit as the perturbation vanishes. It should be noted
that there is no loss in generality here; since there are no additional
degrees of freedom, the freedom in the choice of initial condition
for the perturbed solution $u$ lies in the freedom of choosing the
reference orbit $u_{0}$. If we do a series expansion in eccentricity
$e$ of the right-hand side of eq. \ref{XRef-Equation-32621731},
the equation can be integrated analytically. It turns out that we
can get sufficient accuracy by expanding to third order in eccentricity,
cf. Section \ref{XRef-Subsection-68174859} below. In order to simplify
the expressions and maintain focus on the dominant terms, we will
leave out the higher order terms in the remainder of this section.
The reader may find the corresponding third order expressions in
Appendix \ref{XRef-AppendixSection-68175232}. It should be pointed
out, though, that all perturbative results of the present paper
were derived expanding to third order in eccentricity. The general
solution of eq. \ref{XRef-Equation-32621731} is, to lowest order
in $e$ and $\kappa $:
\begin{multline}
\delta u=\frac{\kappa }{a}-\frac{e}{a}\cos  \nu \\
+A \sin ( \sqrt{1+2 \kappa } \nu ) +B \cos ( \sqrt{1+2\kappa } \nu
) +\mathcal{O}( e^{2}) ,%
\label{XRef-Equation-32771920}
\end{multline}

where $A$ and $B$ are constants of integration. Notice that $\nu
$ is the true anomaly of the reference orbit. Demanding $\delta
u=0$ for $\kappa =0$ fixes $A$ and $B$ to take the values $A=0$
and $B=e/a+\mathcal{O}( e^{2}) $. The perturbation then takes the
following form to lowest order in $e$ and $\kappa $:
\begin{equation}
\delta u=\frac{\kappa }{a}\left( 1-e \nu  \sin  \nu \right) +\mathcal{O}(
e^{2}) .%
\label{XRef-Equation-4483514}
\end{equation}

This gives the perturbation in radial distance, $\delta r$:
\begin{equation}
\delta r=-\frac{\delta u}{u_{0}^{2}}=-\kappa  a( 1 - 2 e \cos  \nu
- e \nu  \sin  \nu ) +\mathcal{O}( e^{2}) .%
\label{XRef-Equation-44162627}
\end{equation}

Thus, the gross effect of a Pioneer type perturbation to the Newtonian
gravitational field is a reduction in the orbital radius of the
order of $ \kappa  a \simeq a \frac{a_{p}}{a_{N}}$. It should be
stated here that the radial perturbation $\delta r$, as given by
eq. \ref{XRef-Equation-44162627}, differs from the expression given
by eq. 58 of Anderson et. al \cite{Anderson:2001sg}. They provide
an expression for $\delta r$ that does not vary with\ \ the longitude
$\varphi $, and therefore can not be a solution to the equation
of motion, eq. \ref{XRef-Equation-32621731}. Eq. \ref{XRef-Equation-44162627}
coincides with eq. 58 of \cite{Anderson:2001sg} in the circular
limit, though. Notice that from eqs. \ref{XRef-Equation-4483514}
and \ref{XRef-Equation-44162627}, we see that $\delta r\sim \kappa
{r_{0}}^{2}/a$. This implies that the more eccentric orbits should
show much stronger variation in the perturbation over the orbital
period than less eccentric orbits. In the circular limit, the perturbation
in orbital radius is constant, ${\delta r}_{e=0}=-\kappa  a$.

Next, let us turn our attention to eq. \ref{XRef-Equation-313225824}.
We can use it to obtain an equation for the perturbation $\delta
\varphi $. Using that $\delta \ell =0$, and that $\frac{\mathrm{d\delta
\varphi }}{\mathrm{dt}}=\frac{\mathrm{d\delta \varphi }}{\mathrm{d\varphi
}}\frac{\mathrm{d\varphi }}{\mathrm{dt}}=\frac{\mathrm{d\delta \varphi
}}{\mathrm{d\varphi }}\frac{\ell }{r_{0}^{2}}$, we get the following
equation for the angular perturbation $\delta$$\varphi$:
\begin{equation}
\frac{d \delta \varphi }{d\varphi  }=2 \frac{\delta u}{u_{0}}%
\label{XRef-Equation-32771848}
\end{equation}

Using eq. \ref{XRef-Equation-4483514}, we get
\begin{equation}
\frac{d \delta \varphi }{d\varphi  }=2 \kappa  \left( 1 - e \cos
\nu  - e \nu  \sin  \nu \right) +\mathcal{O}( e^{2}) .%
\label{XRef-Equation-4495912}
\end{equation}

We can use eq. \ref{XRef-Equation-4495912} to obtain an expression
for the perturbation to the angular velocity $\dot{\varphi }$:
\begin{equation}
\delta \dot{\varphi }=\frac{d \delta \varphi }{d\varphi  }\frac{d\varphi
}{dt}=2 \kappa  \left( 1 - e \cos  \nu  - e \nu  \sin  \nu \right)
\dot{\varphi }+\mathcal{O}( e^{2}) %
\label{XRef-Equation-441104}
\end{equation}

Thus, the angular velocity increases by an amount of the order of
$2\kappa \dot{\varphi }$, which implies a perturbation to the mean
motion of the same order. This is consistent with the conservation
of angular momentum $\ell =r^{2}\dot{\varphi }$, which implies that
$\delta \dot{\varphi }=-2\frac{\delta r}{r}\dot{\varphi }$. As the
planet is drawn closer to the Sun due to an increased gravitational
pull, the angular velocity increases in order for the angular momentum
to remain constant.

Integrating eq. \ref{XRef-Equation-4495912} and demanding that $\delta$$\varphi$
vanishes for $\kappa =0$, we obtain the perturbation $\delta \varphi
$:
\begin{equation}
\delta \varphi =2 \kappa  \left( \nu  + e \nu  \cos  \nu  - 2 e
\sin  \nu \right) +\mathcal{O}( e^{2}) %
\label{XRef-Equation-44105713}
\end{equation}

This is the gross perturbation to the argument of latitude that
should be expected from a Pioneer type perturbation to the Newtonian
gravitational field. The perturbation contains two terms: The first
term is linear in $\nu$, and represents a big secular effect, ${\delta
\varphi }^{(0)}=2 \kappa  \nu $. It comes from an increase in the
mean motion balancing the overall negative perturbation to the orbital
radius. The other term is a smaller ripple on the secular term.

A precise statement should be made regarding the use of the variable
$\varphi$ in the above expressions: $\varphi ( t) $ is the argument
of latitude of the {\itshape reference orbit} at a particular time
$t$ and therefore labels the perturbations uniquely in time:\ \ $\delta
r( t) \equiv \delta r( \varphi ( t) ) $ and $\delta \varphi ( t)
\equiv \delta \varphi ( \varphi ( t) ) $.

\subsection{Accuracy of the perturbative approach}\label{XRef-Subsection-68174859}

A few comments are in place regarding the accuracy of the perturbative
approach used in Section \ref{XRef-Subsection-32891218}: When integrating
eqs. \ref{XRef-Equation-32621731} and \ref{XRef-Equation-4495912}
in order to compute the perturbations $\delta u$ and $\delta \varphi
$, care must be shown when doing series approximations with respect
to the eccentricity of the right-hand sides of the equations. Pluto
has a relatively high eccentricity of $\sim 0.25$. Comparison of
exact numerical with pertubative solutions revealed that we should
expand to 3rd order in eccentricity in order to get sufficient accuracy
in the perturbations to the orbit of Pluto. A comparison of results
from numerical integrations over a complete period of the exact,
non-linear equation of motion, eq. \ref{XRef-Equation-313183311},
with the perturbative Newtonian equations of motion, eq. \ref{XRef-Equation-32621731},
shows that the perturbative solutions for the radial perturbation
$\delta u$ have acceptable accuracy for all three outer planets
when we expand the right-hand side of the equation to third order
in eccentricity. The relative errors $\Delta ( \delta r) /\delta
r$ in the radial perturbation $\delta r$ were $0.02 \%$ and $1.9\%$
for Uranus and Pluto, respectively. Although these are still big
errors in absolute terms due to the large distances involved, they
will not have significant impact on our results, because we are
seeking dominant effects only. A similar comparison of perturbative
and numerical solutions to eq. \ref{XRef-Equation-4495912} shows
that the relative errors $\Delta ( \delta \varphi ) /\delta \varphi
$ were $5\times {10}^{-6}$ and $0.5\%$ for Uranus and Pluto, respectively.\ \ Finally,
the parameter that defines the perturbation to the gravitational
field is the parameter $\kappa $, which is of the order of the ratio
between the anomalous acceleration $a_{p}$ and the mean Newtonian
gravitational acceleration: $\kappa \simeq a_{p}/a_{N}$. $\kappa
$ has a small value for all three planets, of the order of ${10}^{-4}$
for Pluto, and less for the other two. A comparison of exact and
perturbative solutions reveals that a first order perturbative expansion
in $\kappa $ gives good accuracy.\ \

\subsection{Assessment of gross effects for the three outer planets}

\begin{figure}[h]
\begin{center}
\includegraphics{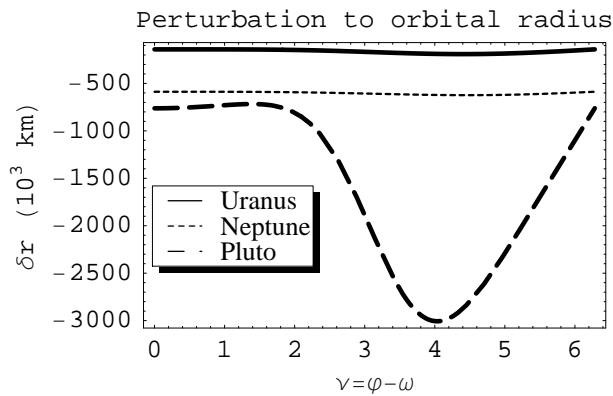}

\end{center}
\caption{Gross perturbation to orbital radius $({10}^{3} \mathrm{km})$
over a complete period for the three outer planets}
\label{XRef-FigureCaption-44163052}
\end{figure}

Let us apply the results of the previous section to the three outer
planets. Eq. \ref{XRef-Equation-44162627} gives the perturbation
$\delta$r to orbital radius as a function of true anomaly $\nu \equiv
\varphi -\omega _{0}$. If we evaluate the $e^{0}$ perturbation ${\delta
r}^{(0)}= -\kappa  a$ for the three outer planets, we get values
of $-155,000$ $ \mathrm{km}$, $-598,000$ $ \mathrm{km}$ and $-1,359,000$
km, respectively, for Uranus, Neptune and Pluto. Figure \ref{XRef-FigureCaption-44163052}
plots $\delta r( \nu ) $\ \ for the three outer planets. As we can
see from the figure, Pluto shows a much stronger variation in $\delta
r$ over the orbital period than the other two planets. This is due
the relative high eccentricity of Pluto's orbit ($\sim 0.25$) and
confirms the strong coupling of $\delta r$ to the orbital radius
mentioned in the previous section.

\begin{figure}[h]
\begin{center}
\includegraphics{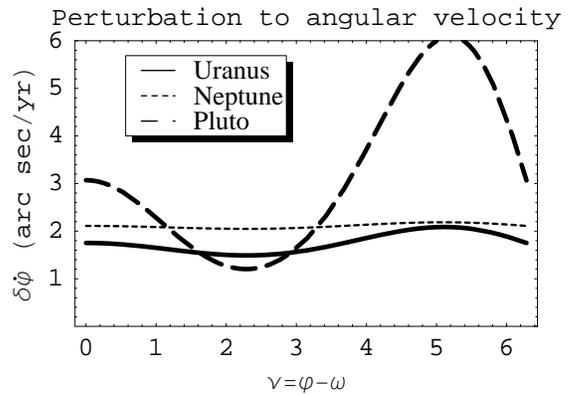}

\end{center}
\caption{Gross perturbation to angular velocity in arc sec/year
over a complete period for the three outer planets}
\label{XRef-FigureCaption-441780}
\end{figure}

The perturbation $\delta \dot{\varphi }$ to the angular velocity
is given by eq. \ref{XRef-Equation-441104}, and the $e^{0}$ perturbation
is $\delta {\dot{\varphi }}^{(0)}=2 \kappa  \dot{\varphi } $. Evaluating
it for the three outer planets gives values of $1.7, 2.1$ and $2.4$
arc sec/year for Uranus, Neptune and Pluto, respectively. $\dot{\varphi
}( \nu ) $ is plotted in Figure \ref{XRef-FigureCaption-441780}.
The variation in the perturbation $\delta \dot{\varphi }$ over the
orbital period is shown in Figure \ref{XRef-FigureCaption-441780}.
As we can see, Pluto has the strongest variation in angular velocity,
which is because of the proportionality $\delta \dot{\varphi }\propto
- \delta r/r^{3}$. $\delta \dot{\varphi }$ increases as $\sqrt{a}$
with orbital radius, and we can clearly see from Figure \ref{XRef-FigureCaption-441780}
that there is stronger variation in the angular velocity perturbation
for orbits with higher eccentricity.

\begin{figure}[h]
\begin{center}
\includegraphics{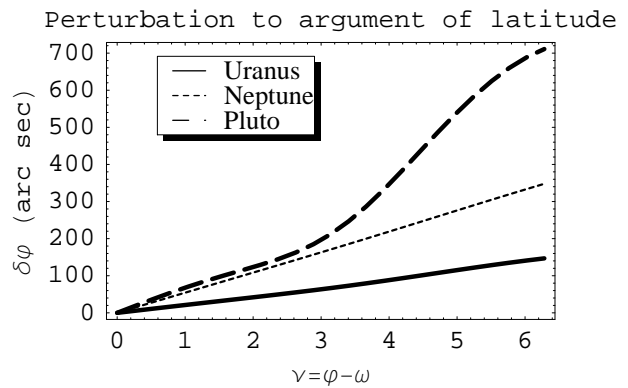}

\end{center}
\caption{Gross perturbation to the argument of latitude $\varphi$
(arc sec) over a complete period for the three outer planets}
\label{XRef-FigureCaption-44183517}
\end{figure}

The perturbation to the argument of latitude, $\delta \varphi $,\ \ is
given by eq. \ref{XRef-Equation-44105713}. The $e^{0}$ term of the
perturbation to the angular velocity $\delta \dot{\varphi }$ implies
a large secular increase of $\delta \varphi $, which is apparent
in the plot of $\delta \varphi $ in Figure \ref{XRef-FigureCaption-44183517}.
For Uranus and Neptune, the secular effect is completely dominant,
and $\delta \varphi $ increases almost linearly with $\varphi$.
For Pluto, however, the big variation in $\delta \dot{\varphi }$
that can be seen in Figure \ref{XRef-FigureCaption-441780} makes
$\delta \varphi $ for Pluto deviate significantly from a linear
increase.\ \ \

\begin{figure}[h]
\begin{center}
\includegraphics{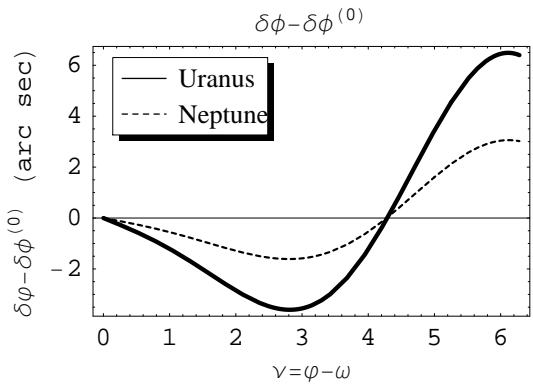}

\end{center}
\caption{The net perturbation to the argument of latitude after
removal of the zero-order secular term}
\label{XRef-FigureCaption-456349}
\end{figure}

If these perturbations were real, one would expect that by fitting
a Keplerian orbit to them that has a slightly higher mean motion,
one would be able to cancel the secular increase. This can easily
be seen by the following example: If the mean motion was increased
by an amount equal to the secular term $\delta {\dot{\varphi }}^{(0)}=2\kappa
\dot{\varphi }$, we would expect the resulting residuals to show
no secular increase. Figure \ref{XRef-FigureCaption-456349} shows
the net perturbation $\delta \varphi -{\delta \varphi }^{(0)}$ for
Uranus and Neptune after removing the secular term ${\delta \varphi
}^{(0)}$. As would be expected, the net perturbations are much smaller,
of the order of a few arc seconds. Notice that, although the gross
$\delta \varphi $ perturbation of Neptune grew much faster than
$\delta \varphi $ for Uranus, due to its higher $\kappa $, the net
perturbations for Neptune is considerably smaller, due to its low
eccentricity.

Later, in Section \ref{XRef-Section-426554}, we will make a thorough
best fit of these $\delta$$\varphi$ perturbations to Keplerian orbits
in order to find the minimal net perturbations for the three outer
planets.

\section{Perturbations to planetary orbits - General Metric Theory
of Gravitation}\label{XRef-Section-415172115}

Now, let us generalize the calculation of the preceeding section
and just assume that the gravitational field can be described by
a metric theory of gravitation. This is a very general case that
includes any PPN-parametrizable theory, including General Relativity.
The purpose of this generalization is to include cases where the
hypothetical perturbation to the gravitational field is not due
to the presence of a gravitational source, but rather is caused
by a deviation in the long-range behavior of the gravitational field
from the behavior expected from General Relativity. As mentioned
in section \ref{XRef-Section-215221244}, alternative gravitational
theories have been proposed as explanations for the Pioneer anomaly,
so we would like our analysis to be able to cover such cases.\ \

\subsection{Deriving the disturbance to the gravitational field}\label{XRef-Section-22165932}

Our first task is to analyze what constraints the introduction of
a constant anomalous radial acceleration would put on the metric
of the surrounding space. We assume a spherically symmetric metric,
and begin by studying the dynamics of a body moving geodesically
in a static, spherically symmetric, helio-centric gravitational
field that can be described by a metric of the form
\begin{equation}
{\mathrm{ds}}^{2}=-B( r) c^{2}{\mathrm{dt}}^{2}+{A( r) }^{-1}{\mathrm{dr}}^{2}+r^{2}(
{\mathrm{d\theta }}^{2}+{\sin }^{2}\theta  {\mathrm{d\varphi }}^{2})
\label{XRef-Equation-115151854}
\end{equation}

An object with rest-mass $\mu  > 0$ moving geodesically in this
gravitational field has two constants of motion; the total energy,
$E$, and the angular momentum, $L$:
\begin{equation}
E=\mu  {c }^{2}B \frac{\mathrm{dt}}{\mathrm{d\tau }}, L= \mu  r^{2}\frac{\mathrm{d\varphi
}}{\mathrm{d\tau }},%
\label{XRef-Equation-115152015}
\end{equation}

where $\tau $ is the proper time of the moving body. The motion
is planar, and we have chosen coordinates so that the plane of motion
coincides with the plane given by $\theta =\pi /2$. Writing the
invariant $g_{\mu \nu }p^{\mu }p^{\nu }=-\mu ^{2}c^{2} $ in the
coordinates of eq. \ref{XRef-Equation-115151854} and applying the
constants of motion of eq. \ref{XRef-Equation-115152015}\ \ yields
the following equations describing geodesic motion in a static,
spherically symmetric gravitational field \cite{Misner-Thorne-Wheeler-1973}:
\begin{gather}
\frac{1}{c^{2}}{\left( \frac{dr}{d\tau }\right) }^{2}=\mathcal{E}^{2}\frac{A}{B}-A(
1+\frac{\ell ^{2}}{c^{2}r^{2}}) %
\label{XRef-Equation-115153055}
\\\frac{d\varphi }{d\tau }=\frac{\ell  }{r^{2}}%
\label{XRef-Equation-24221952}
\\\frac{dt}{d\tau }=\frac{\mathcal{E}}{B}.
\end{gather}

Here we have introduced the constants $\mathcal{E}\equiv E/\mu
c^{2}$ (total energy per unit rest energy) and $\ell \equiv L/\mu
$ (total angular momentum per unit rest mass). Notice that for bodies
moving non-relativistically at speed $v$,\ \ $\mathcal{E}=\sqrt{B(
1+{(v/c)}^{2}) }\simeq 1+\frac{1}{2}{(\frac{v}{c})}^{2}-\frac{G
M}{c^{2}r}$, which is very close to unity. For the three outer planets,\ \ $1-\mathcal{E}=\mathcal{O}(
{10}^{-10}) $.

Notice that these equations are valid for any metric theory of gravitation,
and that we are not making any further assumptions on the\ \ gravitational
interaction besides minimal gravitational coupling, i.e. a body
moving freely in a gravitational field moves geodesically.

In Newtonian gravity, a constant radial acceleration corresponds
to an additional term in the gravitational potential that is linear
in $r$. We then write the metric functions $A$ and $B$ as sums of
the metric function $B_{0}=1-2G M/c^{2}r$ of General Relativity
(representing the unperturbed, expected value of the gravitational
field) and a small perturbation, representing the metric disturbance
necessary to generate the Pioneer anomaly:
\begin{equation}
A( r) \equiv B_{0}( r) +\alpha ( r) , B( r) \equiv B_{0}( r) +\beta
( r) .%
\label{XRef-Equation-22173429}
\end{equation}

The metric ${\mathrm{ds}}_{0}^{2}=-B_{0}c^{2}{\mathrm{dt}}^{2}+B_{0}^{-1}{\mathrm{dr}}^{2}+
r^{2}{\mathrm{d\Omega }}^{2}$ corresponds to a spherically symmetric
vacuum gravitational field around a central body with mass $M$.
We assume $\alpha $ and $\beta $ both to be much smaller than $B_{0}$,
which allows us to expand perturbatively around well-known solutions
for the spherically symmetric vacuum field case. We will refer to
the metric ${\mathrm{ds}}_{0}^{2}$ as the {\itshape unperturbed}
metric. Expressing the radial equation, eq. \ref{XRef-Equation-115153055},
to first order in $\alpha $ and $\beta $, we get
\begin{multline}
\frac{1}{c^{2}}{\left( \frac{\mathrm{dr}}{\mathrm{d\tau }}\right)
}^{2}=\left( \mathcal{E}^{2}+\left( \frac{\mathcal{E}^{2}}{B_{0}}-1\right)
\alpha -\frac{\mathcal{E}^{2}}{B_{0}}\beta \right) \\
-B_{0}( 1+\frac{\ell ^{2}}{c^{2}r^{2}}) %
\label{XRef-Equation-2264043}
\end{multline}

In the following, we restrict our analysis to non-relativistic motion
and weak gravitational fields. In this regime, the $\alpha$-dependence
of eq. \ref{XRef-Equation-2264043} can be discarded, because, with
$v$ being the velocity of the moving object,\ \ $\frac{\mathcal{E}^{2}}{B_{0}}-1={(\frac{v}{c})}^{2}<<1$.
Differentiating eq. \ref{XRef-Equation-2264043} with respect to
proper time $\tau $, we get the expression for the radial acceleration:
\[
\frac{d^{2}r}{{\mathrm{d\tau }}^{2}}=-\frac{1}{2}{c^{2}( B_{0}(
1+\frac{\ell ^{2}}{c^{2}r^{2}}) ) }^{\prime }-\frac{1}{2}{c}^{2}\beta
^{\prime },
\]

where prime denotes differentiation wrt. the radial coordinate $r$.
The first term on the right-hand side of this equation is the expected
radial acceleration generated by the unperturbed vacuum metric.
The last term on the right-hand side can be seen as an anomalous
acceleration produced by an anomalous term $\beta $ in the $g_{00}$
component of the metric. In order for $-\frac{1}{2}{c}^{2}\beta
^{\prime }$ to be a constant, negative acceleration, $\beta$ must
be linear in r:
\begin{equation}
\beta =P r,%
\label{XRef-Equation-27151720}
\end{equation}

with $P$ being a positive constant of dimension $1/r$. It should
be noted that in order to compute the anomalous acceleration observed
by a distant observer, we should transform eq. \ref{XRef-Equation-2264043}
to the proper reference frame of the observer, which for a distant
observer would correspond to coordinate time t at the location of
the observer. However, as $|\mathrm{dt}/\mathrm{d\tau }-1|=\mathcal{O}(
{10}^{-9}) $for the velocities and astronomical distances we are
considering, we can safely equate proper time $\tau$ with coordinate
time $t$. If $a_{P}$ is the anomalous acceleration, directed towards
the Sun, eq. \ref{XRef-Equation-27151720} immediately gives us the
relationship between the metric parameter\ \ $P$ and $a_{P}$: $P
= 2 a_{P}/c^{2}\mathcal{E}^{2}$.

We can conclude that an observed constant anomalous acceleration
uniquely constrains the dominant part of the disturbance to the
gravitational potential. An additional term, linear in $r$, must
be present in the $g_{00}$ component of the metric in order to produce
such an effect. This result is the same we get from Newtonian gravity.
The analysis is independent of how this metric disturbance is produced,
and is therefore also independent of the particular gravitational
theory applied to analyze the the disturbance.

\subsection{Equations of motion}\label{XRef-Section-2655810}

Having found the perturbation to the gravitational field, we are
now ready to explore the equations of motion for a body moving geodesically
in this field. When discarding the $\alpha$-term of eq. \ref{XRef-Equation-2264043}
and using that $\frac{\mathcal{E}^{2}}{B_{0}}\beta  \simeq  \beta
$, we\ \ get the radial equation of motion
\begin{equation}
\frac{1}{c^{2}}{\left( \frac{\mathrm{dr}}{\mathrm{d\tau }}\right)
}^{2}=\left( \mathcal{E}^{2}-\beta \right) -B_{0}( 1+\frac{\ell
^{2}}{c^{2}r^{2}})
\end{equation}

By using eq. \ref{XRef-Equation-24221952},\ \ expressing the equation
in terms of the reciprocal radial coordinate $u\equiv 1/r$,\ \ taking
the derivative of the equation with respect to $\varphi$ and using
eq. \ref{XRef-Equation-27151720}, we get the following\ \ second-order
equation for $u( \varphi ) $:\ \
\begin{equation}
\frac{d^{2}u}{{\mathrm{d\varphi }}^{2}}+\left( 1-\frac{3G M }{c^{2}}u\right)
u=\frac{G M }{\ell ^{2}}+\frac{ a_{P}}{\ell ^{2}\mathcal{E}^{2}}\frac{1}{u^{2}}%
\label{XRef-Equation-4112624}
\end{equation}

The unperturbed $(a_{p}=0)\text{}$ equation has the following approximate
solution valid for low eccentricity orbits \cite{Misner-Thorne-Wheeler-1973,Carroll-2004}:
\begin{equation}
\frac{1}{u_{0}}=r_{0}=\frac{a( 1-e^{2}) }{1+e \cos  \left( f \left(
\varphi -\omega _{0}\right) \right) },%
\label{XRef-Equation-4112552}
\end{equation}

where $e$ is the eccentricity of the orbit, $\omega _{0}$ is the
argument of the perihelion at a time when the planet is at perihelion,
and $f^{2}=1-6{(\frac{G M}{c \ell })}^{2}\rightarrow 1$ as $r_{0}\rightarrow
\infty $. $f$ is very close to unity. When setting $f=1$ we get
the closed Keplerian orbits of eq. \ref{XRef-Equation-32618372}.
In general, a factor $f < 1$ yields a small relativistic perihelion
precession, which for low eccentricity orbits is $\simeq 6\pi  G
M/(c^{2}a( 1-e^{2}) )$\ \ per cycle. Evaluating the relativistic
perihelion precession for the three outer planets gives values of
of the order of ${10}^{-3}$ arc second per cycle. This is well below
what would be observable, given the present data and their accuracy,
which is of the order of 0.5 arc second.

Let us proceed as we did for for the Newtonian case of Section \ref{XRef-Subsection-32891218}\ \ and
assume that $u_{0}$ is a reference solution to the unperturbed equations
of motion, of the form given by eq. \ref{XRef-Equation-4112552}.
Let us then define the perturbation $\delta u=u-u_{0}$, where $u$
is a solution to the perturbed equation of motion, eq. \ref{XRef-Equation-4112624},
having the same angular momentum $ \ell $ as $u_{0}$. We obtain
the following equation of motion for the perturbation $\delta u$:
\begin{equation}
\frac{d^{2}\delta u}{{\mathrm{d\varphi }}^{2}}+\left( f^{2}+2\kappa
\right) \delta u=\frac{\kappa }{a( 1-e^{2}) {\left( 1+e \cos  \left(
f \nu \right) \right) }^{2}},%
\label{XRef-Equation-41121419}
\end{equation}

where $\nu \equiv \varphi -\omega _{0}, \kappa \equiv \frac{a_{p}}{a_{N}
}{(1-e^{2})}^{2}, a_{N}\equiv \frac{G M}{a^{2}}$, as in the Newtonian
case before. Now, $ 1-f^{2}$ is of the order ${10}^{-9}$ for the
three outer planets. Since $\kappa $ is $\mathcal{O}( {10}^{-4})
$, we can safely ignore any Post-Newtonian corrections to the solutions,
set $f=1$ and proceed with the Newtonian equation of motion, eq.
\ref{XRef-Equation-32621731} and the corresponding solutions, as
derived in section \ref{XRef-Section-4264724}.

\section{Analysis Method}\label{XRef-Section-4264518}

\subsection{Observations and their residuals}\label{XRef-Subsection-312182718}

The procedures that are applied for calculating the ephemerides
of the planets use available theory, taking into account known gravitational
sources, and carefully fitting the computations to available observational
data \cite{Standish-1998,Pitjeva-2005a}. One would therefore expect
that any disturbance to the gravitational field within the solar
system would, if large enough, over time generate residuals in the
position measurements of the planets as they would tend to drift
from their expected positions. The question then is, what residuals
would arise if we were to explain the Pioneer anomaly as the result
of a perturbation to the gravitational field in the outer solar
system, and how do they compare with the observed residuals? In
this section, we will elaborate the analysis method which we will
apply in order to answer these questions.

Before we start our analysis, let us review what the observables
are and how to compare models with observations. See Appendix \ref{XRef-AppendixSection-31115539}
for a review of terminology and notation used. When restricted to
performing optical observations only of a distant object, which
has been the case for the three outer planets, there are just two
observables: the {\itshape right ascension (r.a.)},\ \ and {\itshape
declination (dec.)}. These are the position angles that locate the
object on the celestial sphere. Accurate optical measurements of
the positions of the outer planets have been performed since the
early 20th century. The a priori uncertainty of these measurements
were $0.5-1$ arc second during most of this period \cite{Pitjeva-2005a}.
In recent years, the a priori accuracy has been improved to $~0.2$
arc second. Computational models of the motion of the known objects
in the Solar System can then be applied to compute the ephemerides,
which are the expected orbits, of these objects. The free parameters
of the model are fixed by making a best fit of the computations
with the observations. The results are the Solar System Ephemerides,
composed of highly accurate maps of the motions of the planets as
well as other known Solar System bodies \cite{Pitjeva-2005a,Standish-1998}.
\begin{figure}[h]
\begin{center}
\includegraphics{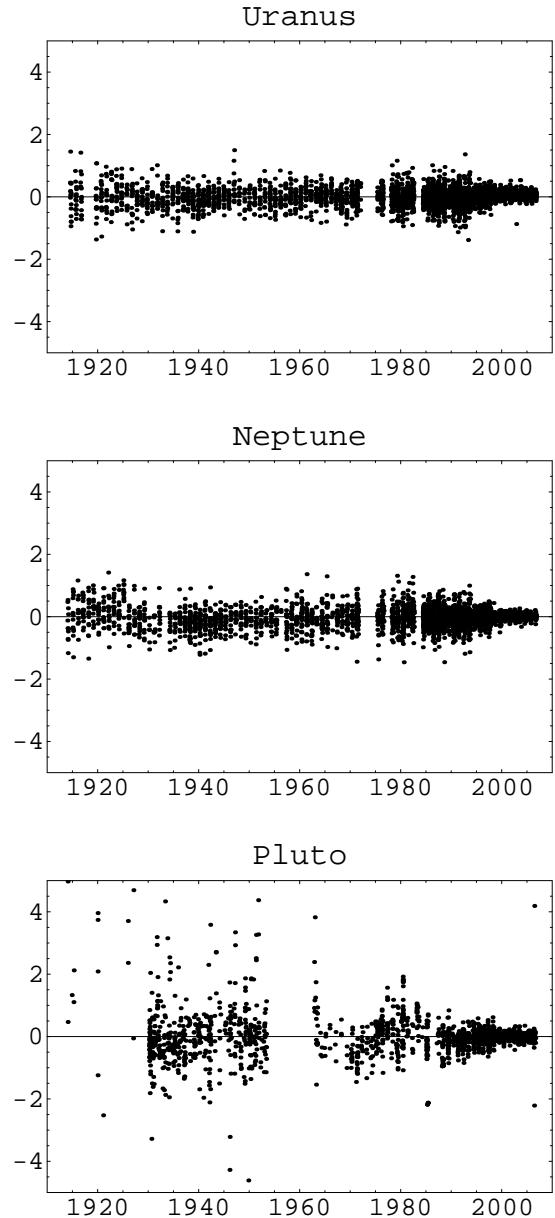}

\end{center}
\caption{Actual r.a. residuals (in arc seconds) for the three outer
planets, courtesy of E.M. Standish, JPL}
\label{XRef-FigureCaption-311224056}
\end{figure}

Figure \ref{XRef-FigureCaption-311224056} shows the DE414 residuals\cite{Standish-1998}
in right ascension for the three outer planets. These residuals
are the difference between the measured and ephemeris right ascension
values, taken over the entire span of available data. As could be
expected, the residuals are smaller towards the end of the data
span, apparently due to the increased accuracy of the measurement
techniques applied. Table \ref{XRef-TableTitle-311224934} gives
the standard deviations of the r.a. residuals for the three outer
planets\footnote{For the Pluto data set, a few data points with
ra or dec of 10 arc seconds or more were deleted. These residuals
were more than 10 times the a priori measurement accuracy, and were
therefore deemed to be due to invalid measurements. }. The standard
deviations of these residuals are within the a priori measurement
accuracies that have been reported \cite{Standish-1998,Pitjeva-2005a}.
\begin{table}
\caption{Standard deviations in right ascension residuals for the
three outer planets} \label{XRef-TableTitle-311224934}
\begin{ruledtabular}

\begin{tabular}{lc}
Planet & Std. dev - $\sigma _{l}$

(arc sec.)\\
\hline
Uranus & 0.3\\
Neptune & 0.3\\
Pluto & 0.8
\end{tabular}
\end{ruledtabular}
\end{table}

\subsection{Description of the analysis method}\label{XRef-Subsection-32785054}

Our best option for determining whether or not a perturbation to
the gravitational field would have an observable effect on the motions
of the outer planets is to solve the basic equations of motion for
the three planets under the hypothesis of a perturbed gravitational
field. The purpose of these solutions,\ \ referred to as the {\itshape
simulated observations orbits}, is to provide targets for best fit
approximations using solutions to the unperturbed equations of motion.
Just like real ephemerides emerge as best fits of solutions of the
equations of motion to the real observations, we will construct
{\itshape simulated ephemerides} by fitting solutions of the unperturbed
equations of motion\ \ to the simulated observations orbits\footnote{We
fit solutions to the unperturbed equations of motion to solutions
of the perturbed equations of motion. This is a simplification compared
to the way the real ephemeris is constructed by fitting to real
observations. However, we expect this simplification to have no
impact on the end result, i.e. the best fit solution will provide
the best fit regardless of the reference frame in which the fit
was made}. The simulated ephemerides obtained this way therefore
play the same role as the real ephemerides, i.e. the best fit of
known models to observations. From the simulated observations orbit
and the corresponding simulated ephemeris, we can derive the corresponding
values of the optical observables; the right ascension $\alpha $
and the declination $\delta $ \footnote{The reason why we choose
to work from the basic equations of motion and not to use a perturbative
approach like the Gauss' equations is to be able to generalize the
approach to any metric theory of gravity.} and their residuals.
These residuals (simulated obs. value - simulated ephemeris value),
which are derived from the basic equations of motion, provide a
realistic lowest order estimate of the residuals in the observables
to be expected from a perturbation to the gravitatational field
of the kind required to account for the Pioneer anomaly.\ \

Although we work with models that are idealized compared to the
realistic models underlying modern planetary ephemerides, we still
expect the simulated residual estimates to be realistic to lowest
order in the perturbations, because to lowest order, perturbations
decouple and can be studied independently.

Finally, since we would like to compare the results of the hypothetical
model to real observations, we will define the {\itshape predicted
observations} to be the sum of the real ephemeris values and the
best fit simulated residuals. The underlying assumption here is
that the simulated residuals can be used to predict the offset between
the ephemeris and the expected observations under the hypothesis
of a perturbed gravitational field. This is the closest we can get
with respect to checking this hypothesis against data. We are not
interested in the predicted observations themselves, only in their
residual relative to the real observations. As we will see later,
the (real -predicted) residual can be estimated by combining the
real and simulated residuals.

Having obtained the simulated residuals as the difference between
the simulated observations and the simulated ephemerides, the remaining
problem is how to compare them with the real residuals, described
in section \ref{XRef-Subsection-312182718} above. Let us formulate
two hypotheses that we would like to test by statistical methods:

{\bfseries H1}: Given the uncertainty of the observations, the {\itshape
simulated r.a. observation}s do not deviate significantly from the
{\itshape simulated ephemerides}

{\bfseries H2}: Given the uncertainty of the observations, the {\itshape
predicted r.a. observations} do not deviate significantly from the
{\itshape real observations}

Hypothesis H1 is testing, under the assumption of a perturbed gravitational
field, whether the ephemerides still would provide good fits to
the observations. If not, it would contradict the present state
of matters, because the ephmerides do provide good fits to the data.

Hypothesis H2 tests, again under the assumption of a perturbed gravitational
field, whether the model predictions are consistent with the actual
observations.

Now, let us restate the original hypothesis stated in Section \ref{XRef-Subsection-41364936}
above:

{\bfseries H0}: The Pioneers move geodesically in a perturbed gravitational
field

The logical connection between hypothesis H0 and the other two should
be clear. H0 is falsified if any of hypotheses H1 and H2 are falsified,
because falsification of H1 or H2 would imply that a perturbation
to the gravitational field required to induce the Pioneer anomaly
is not present. On the other hand, H1 and H2 can obviously not be
used to verify H0.

Notice that the results will critically depend on judicial estimates
of the measurement uncertainties. For a given planet $l$, we will
use the standard deviation $\sigma _{l}$ of the r.a. residuals shown
in Figure \ref{XRef-FigureCaption-311224056} as the best estimate
of the uncertainty in the observations. If the ephemeris provides
a good match to the observations, it is reasonable to assume that
$\sigma _{l}$ will be close to the measurement uncertainty. If not,
$\sigma _{l}$ would be\ \ higher than the measurement uncertainty.
In either case, using $\sigma _{l}$ as a best estimate of the uncertainty
in the measurements seems a reasonable, if not conservative, choice.

In order to test hypotheses H1 and H2, we will perform a chi square
analysis with a confidence limit of 99\%. Let us elaborate how we
will apply this method in our case. Given $N_{l}$ measurements $({\hat{\alpha
}}_{l,j},t_{j})$ of the right ascension for planet $l$ at measurement
times $t_{j}$, the real residuals $\rho _{l, j}( t_{j}) $ for planet
$l$ is the difference between the observation ${\hat{\alpha }}_{l,j}$
and the corresponding ephemeris value\ \ $\alpha _{l}( t_{j}) $:\ \ $\rho
_{l, j}( t_{j}) \equiv {\hat{\alpha }}_{l,j}-\alpha _{l}( t_{j})
$. These are the residuals that are plotted in Figure \ref{XRef-FigureCaption-311224056}
above, which we will refer to as the {\itshape real residuals}.

Given a solution to the equations of motion for planet $l$ under
the hypothesis that the gravitational field is perturbed by a constant
radial acceleration $a_{p}$ directed towards the Sun, we can derive
the set of simulated right ascension values ${\hat{\alpha }}_{l}^{P}(
t_{j}) $. We will refer to these as our set of {\itshape simulated
observations}. The {\itshape simulated ephemeris} is the solution
of the unperturbed equations of motion that provides the best fit
to the simulated observations orbits, and from it we can derive
a set of simulated right ascension ephemeris values $\alpha _{l}^{P}(
t_{j}) $. The {\itshape simulated residuals} $\rho _{l,j}^{P}$ are
then the difference between the simulated right ascension values
and the corresponding simulated ephemeris values: $\rho _{l, j}^{P}(
t_{j}) \equiv {\hat{\alpha }}_{l,j}^{P}( t_{j}) -\alpha _{l}^{P}(
t_{j}) $.

The {\itshape predicted} r.a. value ${\hat{\alpha }}_{l,j}^{P^{\prime
}}( t_{j}) $ is the sum of the real ephemeris value $\alpha _{l}(
t_{j}) $ and the simulated residual $\rho _{l,j}^{P}$:
\begin{equation}
{\hat{\alpha }}_{l,j}^{P^{\prime }}( t_{j}) \equiv \alpha _{l}(
t_{j}) +\rho _{l,j}^{P}
\end{equation}

Define the real-predicted observation residual\ \ $\rho _{l,j}^{P^{\prime
}}$ to be the difference between the real observation and the predicted
observation:
\begin{equation}
\rho _{l,j}^{P^{\prime }}\equiv {\hat{\alpha }}_{l,j} \left( t_{j}\right)
-{\hat{\alpha }}_{l,j}^{P^{\prime }}( t_{j}) =\rho _{l, j}( t_{j})
-\rho _{l,j}^{P}( t_{j}) %
\label{XRef-Equation-4117401}
\end{equation}

We see from eq. \ref{XRef-Equation-4117401} that we can estimate
the real-predicted observation residuals by combining the real residuals
$\rho _{l, j}( t_{j}) $ with the simulated residuals $\rho _{l,j}^{P}(
t_{j}) $. This allows us to compare the predicted observations with
the real observations using the chi square method.

Let $\sigma _{l}^{2}$ be the variance of the real residuals for
planet $l$. As discussed previously, we will assume that $\sigma
_{l}$ is a good estimate of the measurement uncertainty. Now, define
the following chi square values that we will use to test the hypotheses
above:
\begin{gather}
\chi _{l,1}^{2}\equiv \frac{1}{\sigma _{l}^{2}}\sum \limits_{j=1}^{N_{l}}\rho
_{l,j}^{P} {\left( t_{j}\right) }^{2}
\\\chi _{l,2}^{2}\equiv \frac{1}{\sigma _{l}^{2}}\sum \limits_{j=1}^{N_{l}}{\left(
\rho _{l,j}^{P^{\prime }}\right) }^{2}
\end{gather}

Provided measurement errors are distributed normally, a model deviates
from observations with probability $p$ if the $\chi ^{2}$ of the
residuals between the observations and the model prediction exceeds
$N_{l}+C_{p}$, where $C_{p}$ is the confidence limit applied. $C_{p}$
depends on the probability given (the confidence level) and the
number of free parameters. In our case, the number of free parameters
is 4, and we have chosen a confidence level of 0.99. This gives
$C_{p}=13$.

$\chi _{l,1}^{2}$ can be used to test hypothesis H1 above; if $\chi
_{l,1}^{2}>N_{l}+C_{p} $, the deviation of the simulated r.a. observations
from the simulated ephemeris is statistically significant at a significance
level of $p$. Similarly,\ \ $\chi _{l,2}^{2}$ can be used to test
hypothesis H2 above; if $\chi _{l,2}^{2}>N_{l}+C_{p} $, the deviation
of the projected r.a. observations from the real observations is
statistically significant at a significance level of $p$.

\subsection{Comments on the perturbative approach for solving the
equations of motion}

In this paper, we study what observational effects a disturbance
to the gravitational field would have on orbits of a planet moving
in a static, spherically symmetric gravitational field around the
Sun. We work perturbatively to lowest order in the perturbation.
Since we are looking for dominant effects only, we are free to disregard
other perturbing effects, such as the presence of other bodies in
the Solar System. These simplifications allow us to treat the problem
analytically.

In section \ref{XRef-Section-4264724}, we started our analysis by
choosing a reference orbit, which is an idealized orbit satisfying
the unperturbed equations of motion and matching the orbital parameters
of one of the three outer planets. In the Newtonian case, the reference
orbit is a Keplerian orbit that is set up by applying the mean Keplerian
elements of the planet as initial conditions. This reference orbit
has no other role than being the reference for two perturbations:

In section \ref{XRef-Section-4264724}, we found the first pertubation,
which gives rise to the simulated observations. It comes from solving
the equations of motion for a perturbed gravitational field. For
this orbit, the {\itshape simulated observations orbit}, we applied
the assumption that the gravitational field is perturbed in a way
that results in a constant, anomalous radial acceleration $a_{P}\equiv
8.7\times {10}^{-10}m/s$ towards the Sun. By integrating the equations
of motion, we obtained the perturbation to the argument of latitude,
denoted by $\delta \varphi $. The argument of latitude, $\varphi$,
is related to the true anomaly, $\nu$, by $\varphi \equiv \omega
+\nu $, where $\omega $ is the argument of periapsis \footnote{The
reason for using the argument of latitude as a model parameter instead
of the true anomaly is just formal; that the former is a coordinate
that is relative to a point that in our idealized models is fixed
(the ascending node, i.e. the point where the object ascends through
the ecliptic), whereas the true anomaly is measured relative to
the a point that moves; the periapsis.}. In our case, $\omega$ is
the argument of periapsis of the reference orbit, i.e. a fixed position.
If $\varphi _{0}( t) $ is the argument of latitude of the reference
orbit, $\varphi _{p}( t) =\varphi _{0}( t) +\delta \varphi ( t)
$ is the argument of latitude of the perturbed orbit. Thus, we can
derive expected values for the $\alpha$ and $\delta$ observables
from $\varphi _{p}$ by performing a coordinate transformation of
the position\ \ vector $(r_{p},\varphi _{p})$ in the orbital plane
of the planet to geocentric equatorial coordinates.\ \

The second perturbation, which we will refer to as our {\itshape
simulated ephemeris},\ \ arises when we attempt to find the solution
that best fits the simulated observations orbit $(r_{p},\varphi
_{p})$. In order to simplify the analytical treatment, we assume
that this solution can be treated as a perturbation, denoted $\Delta
\varphi $, to the reference orbit as well. Having found the best
fit solution by performing a least squares approximation of the
second perturbed orbit $(r_{E}( t) =r_{0}( t) +\Delta r( t) ,\varphi
_{E}( t) =\varphi _{0}( t) +\Delta \varphi ( t) )$ to the simulated
observations orbit $(r_{p},\varphi _{p})$, we can estimate the simulated
residuals in $\alpha$ and $\delta$ as the difference between the
values of the simulated observations and the simulated ephemeris,
as described in section \ref{XRef-Subsection-32785054}.

\section{Residual Analysis}\label{XRef-Section-426554}

If $(r_{0}( \varphi _{0}) , \varphi _{0}( t) )$ is the reference
orbit, $(r_{p}( \varphi ) =r_{0}( \varphi _{0}) +\delta r( \varphi
_{0}) , \varphi _{p}( t) =\varphi _{0}( t) +\delta \varphi ( \varphi
_{0}) )$ is the simulated observations orbit, i.e. the perturbed
orbit from which we can derive simulated observations. The simulated
observations orbit was derived in section \ref{XRef-Section-4264724}.
Now it is time to proceed with the next step of our analysis: Following
the analysis procedure outlined in Section \ref{XRef-Subsection-32785054}
above, our next task is to compute the simulated ephemeris, which
is the\ \ Keplerian orbit that provides the best fit to the simulated
observations orbit.

\subsection{Finding the simulated ephemerides}\label{XRef-Subsection-61013050}

Eq. \ref{XRef-Equation-313195439} represents the unperturbed equation
of motion that we will solve. Since the simulated observations orbit
$(r_{p}( \varphi ) , \varphi _{p}( t) )$ is just a perturbation
of the reference orbit $(r_{0},\varphi _{0})$, we can assume that
the best fit Keplerian orbit, our simulated ephemeris, also represents
a pertubation to the reference orbit. Perturbing eq. \ref{XRef-Equation-313195439}
gives the equation for the radial perturbation $\Delta u\equiv u-u_{0}$,
where $u$\ \ here is another Keplerian orbit with angular momentum
$\ell \equiv \ell _{0}+\Delta \ell $:
\begin{equation}
\frac{d^{2}\Delta u}{{\mathrm{d\varphi }}^{2}}+\Delta u=-\frac{2}{a(
1-e^{2}) }\frac{\Delta \ell }{\ell _{0}}.%
\label{XRef-Equation-327215430}
\end{equation}

Eq. \ref{XRef-Equation-327215430} has the general solution
\begin{equation}
\Delta u=\frac{1}{a}\left(  e C_{1}\sin \ \ \nu + e C_{2}\cos  \nu
-\frac{2}{\left( 1-e^{2}\right) }\frac{\Delta \ell }{\ell _{0}}\right)
,%
\label{XRef-Equation-49171555}
\end{equation}

where $C_{1}$, $C_{2}$ and $\Delta$$ \ell $ will be treated as free
parameters. Similarly to how we derived eq. \ref{XRef-Equation-32771848},
we get the equation for the angular perturbation:
\begin{equation}
\frac{d \Delta \varphi }{\mathrm{d\varphi } }=\frac{\Delta \ell
}{\ell _{0}}-2\frac{\Delta r}{r_{0}}%
\label{XRef-Equation-327221251}
\end{equation}

The exact solution to eq. \ref{XRef-Equation-327221251} is
\begin{widetext}\begin{equation}
\Delta \varphi =C_{0}+\left( \frac{\Delta \ell }{\ell } +2 C_{2}
\left( 1-e^{2}\right)  \right)  \nu -2 C_{1}( 1-e^{2})  \ln ( 1+e
\cos  \nu ) -\frac{4}{\sqrt{1-e^{2}}} \left( 2 \frac{\Delta \ell
}{\ell }+ C_{2}( 1-e^{2})  \right)  \arctan ( \sqrt{\frac{1-e}{1+e}}
\tan \frac{\nu }{2})
\end{equation}
\end{widetext}

Define $u_{K}\equiv u_{0}+\Delta u$ and\ \ $\varphi _{K}\equiv \varphi
_{0}+\Delta \varphi $. $\Delta \varphi $ introduces another free
parameter $C_{0}$. This leaves us with four free parameters that
we will fit to the simulated orbit: $C_{0}$, $C_{1}$, $C_{2}$ and
$\Delta$$ \ell $.\ \

Define the simulated residual $\hat{\Delta \varphi }$ by\ \ $\hat{\Delta
\varphi }\equiv \varphi _{p}( \varphi _{0}) -\varphi _{K}( \varphi
_{0}) =\delta \varphi -\Delta \varphi $. The simulated residual
$\hat{\Delta \varphi }$\ \ can be written in the form
\begin{equation}
\hat{\Delta \varphi }=\delta \varphi -C_{0}+\frac{\Delta \ell }{\ell
} f_{0}( \varphi ) +C_{1}f_{1}( \varphi ) + C_{2}f_{2}( \varphi
) ,%
\label{XRef-Equation-6913130}
\end{equation}

where $f_{0}, f_{1}$ and $f_{2}$ are functions that to third order
in eccentricity take the form
\begin{multline*}
f_{0}\equiv  \left( 3+2 e^{2}\right)  \nu - e \left( 4+3 e^{2}\right)
\sin  \nu +e^{2} \sin  2 \nu \\
-\frac{e^{3}}{3} \sin  3 \nu +\mathcal{O}( e^{4})
\end{multline*}
\begin{multline*}
f_{1}\equiv e \left\{ -\frac{e}{2}\ \ + \left( 2-\frac{3 }{2}e^{2}\right)
\cos  \nu -\frac{e}{2}\cos  2\nu +\frac{e^{2}}{6} \cos  3\nu \right\}
\\
+\mathcal{O}( e^{4})
\end{multline*}
\begin{multline*}
f_{2}\equiv e\left\{  e \nu - \left( 2-\frac{e^{2}}{2}\right)  \sin
\nu +\frac{e }{2} \sin  2\nu -\frac{e^{2}}{6} \sin  3\nu \right\}
\\
+\mathcal{O}( e^{4})
\end{multline*}

Let us consider for a moment the circular limits of eqs. \ref{XRef-Equation-49171555}
and \ref{XRef-Equation-6913130}.\ \ In that case, $f_{1}=f_{2}=0$.
Eqs. \ref{XRef-Equation-4483514} and \ref{XRef-Equation-49171555}
show that in the circular limit, the simulated radial residual $\delta
u-\Delta u$ can be eliminated by properly adjusting the angular
momentum of the the Keplerian orbit. However, in this case we are
left with a non-vanishing simulated angular residual $\hat{\Delta
\varphi }$. If we, on the other hand, adjust the angular momentum
and $C_{0}$ to make the angular residual $\hat{\Delta \varphi }$
vanish, we are left with a non-zero radial residual $\hat{\Delta
r}\equiv \delta r-\Delta r=\kappa  a/3$. Thus, even in the circular
limit, we can not simultaneously fit both the angular and radial
perturbations by a Keplerian orbit. When projecting $\hat{\Delta
r}$ onto the celestial sphere for the near circular orbits of Uranus
and Neptune, it translates to perturbations in the observation angles
of $0.2"$ and $0.3"$, respectively. This implies that in the circular
limit, even if the angular residual $\hat{\Delta \varphi }$ can
be eliminated by adjusting the angular momentum, the Pioneer effect
is in principle still observable from the perturbation to the observation
angle resulting from the non-vanishing radial residual $\hat{\Delta
r}$. In the following, we will make a best fit of the simulated
angular residual $\hat{\Delta \varphi }$ and ignore the effect of
the remaining radial perturbation on the observable angle.

Now, let us define the best fit solution $(r_{E}, \varphi _{E})$
as the solution $(r_{K},\varphi _{K})$ that minimizes the variance
of the simulated residual $\hat{\Delta \varphi }$ over the measurement
interval [$t_{0}, t_{1}$]:\ \
\begin{equation}
\sigma _{p}^{2}\equiv  \frac{1}{\left( \varphi ( t_{1}) -\varphi
( t_{0}) \right) }\operatorname*{\overset{\varphi ( t_{1}) }{\int
}}\limits_{\varphi ( t_{0}) }d\varphi \ \ {\hat{\Delta \varphi }}^{2}%
\label{XRef-Equation-3272356}
\end{equation}

 The best fit is made by finding the values of the free parameters
$\Delta \ell $, $C_{0}$, $C_{1}$ and $C_{2}$ that give minimal variance
$\sigma _{p}^{2}$. Equation \ref{XRef-Equation-3272356} can be integrated
analytically, and gives a second order polynomial in the free parameters
$C_{0}$, $C_{1}$, $C_{2}$ and $\Delta$$ \ell $. This polynomial
has a unique minimum that can be found by solving the equations
$\frac{\partial }{\partial C_{0}}\sigma _{p}^{2}=0$ , $\frac{\partial
}{\partial C_{1}}\sigma _{p}^{2}=0$, $\frac{\partial }{\partial
C_{2}}\sigma _{p}^{2}=0$ and $\frac{\partial }{\partial \Delta \ell
}\sigma _{p}^{2}=0$.

\subsection{Analyzing the simulated residuals}

\begin{figure}[h]
\begin{center}
\includegraphics{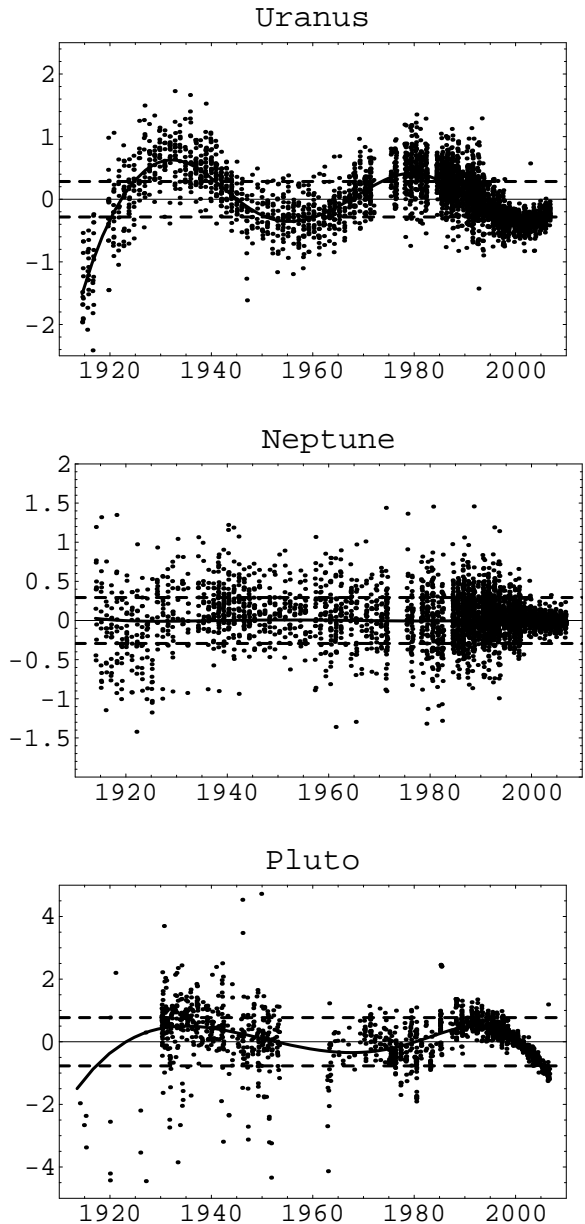}

\end{center}
\caption{Simulated (solid lines) and predicted-real r.a. residuals
(in arc seconds) for the three outer planets. Dashed lines indicate
the standard deviation of the real residuals.}
\label{XRef-FigureCaption-328101558}
\end{figure}

Figure \ref{XRef-FigureCaption-328101558} plots $-\rho ^{p^{\prime
}}$, the predicted-real r.a. residuals, and the simulated r.a. residuals
for the three outer planets. In order to make the simulated residuals
easy to compare with the real residuals, a coordinate transformation
of the orbit to the mean equatorial plane has been done. The plots
also show the standard deviations of the real residuals.\ \

From Figure \ref{XRef-FigureCaption-328101558} we can immediately
see that the simulated residuals for Neptune are very small compared
to the real residuals. This implies that a Pioneer-type perturbation
would have no observable effect on the orbit of Neptune. Furthermore,
by visual inspection of the real-predicted residuals in\ \ Figure
\ref{XRef-FigureCaption-328101558}, we see that there are significant
deviations between real and predicted observations for Uranus and
Pluto.

Now, let us put this on formal grounds and test hypotheses H1 and
H2 formulated in Section \ref{XRef-Subsection-32785054} above. Given
the simulated residuals $\rho ^{p}$ and the real residuals $\rho$,
we must compute the chi square values $\chi _{l,1}^{2}$\ \ and $\chi
_{l,2}^{2}$ that target these hypotheses. Table \ref{XRef-TableTitle-32963136}
shows the results of this computation and the tests of hypotheses
H1 and H2 for each of the three outer planets.\ \
\begin{table}
\caption{Chi square values of the simulated residuals for the three
outer planets and the results of the tests of hypotheses H1 and
H2 of Section \ref{XRef-Subsection-32785054}} \label{XRef-TableTitle-32963136}
\begin{ruledtabular}

\begin{tabular}{lcccccc}
Planet & $\sigma _{l} (")$ & $N_{l}$ & $\chi _{l,1}^{2}-N_{l}$ &
$\chi _{l,2}^{2}-N_{l}$ & H1 & H2\\
\hline
Uranus & 0.283 & 3678 & 486 & 4930 & False & False\\
Neptune & 0.293 & 3800 & -3799 & 6.5 & True & True\\
Pluto & 0.771 & 2119 & -1510 & 773 & True & False
\end{tabular}
\end{ruledtabular}
\end{table}

The main conclusions drawn from this analysis are that, given an
uncertainty $\sigma _{l}$ in the r.a. observations:
\begin{enumerate}
\item For Uranus, the simulated r.a. observations show a deviation
from the simulated ephemeris that is statististically significant
well beyond the 99\% confidence level. For Pluto, however, no such
deviation is found for the simulated r.a. observations. Thus, H1
is falsified for Uranus and verified for the other two planets.
\item For both Uranus and Pluto, the predicted r.a. observations
show a deviation from the real r.a. observations that is statististically
significant well beyond the 99\% confidence level. Thus, H2 is falsified
for Uranus and Pluto, but verified for Neptune.
\end{enumerate}

\section{Conclusions}\label{XRef-Section-215221331}

Under the assumption that the observed frequency drift of the Pioneer
spacecraft is caused by a constant, anomalous acceleration that
is directed towards the Sun, and that the Pioneer spacecraft move
geodesically in a perturbed gravitational field that is static and
spherically symmetric, we have analyzed the effects that such a
perturbation to the gravitational field would have on the orbits
of the three outer planets. The basic assumptions are that the gravitational
field can be modeled within a metric theory of gravity and that
particles couple minimally to the gravitational field, i.e. they
move geodesically when falling freely in a gravitational field.
Our results are model independent within this class of theories
in the sense that they do not depend on the particulars of the gravitational
theory, e.g. the nature of any gravitational source or how gravitational
sources couple to the spacetime metric. The analysis is valid in
the weak field limit of any metric gravitational theory, and it
therefore applies to any gravitational theory that can be parametrized
within the PPN framework.

We showed in section \ref{XRef-Section-22165932} that the metric
perturbation that would be needed to induce the Pioneer anomaly
can be uniquely constrained to lowest order. This immediately implies
that perturbed orbits due to the perturbation in the gravitational
field will be the same, to lowest order in the perturbation, for
all metric theories of gravity. Furthermore, by solving the basic
equations of motion for a perturbed gravitational field, we indeed
found that Post-Newtonian effects were small compared to Newtonian
effects.

By solving the unperturbed equations of motion and finding the solutions
that provide the best fit to the perturbed solutions, we were able
to derive simulated right ascension residuals that could be compared
with the real ephemeris residuals.

We then formulated an analysis method, based on the chi square method,
for comparing simulated residuals with the real ephemeris residuals.
We formulated two hypotheses that, if falsified, would falsify the
hypothesis of the presence of a perturbation to the gravitational
field. Following from Einstein's equivalence principle, this would
imply that the Pioneers do not move geodesically in a perturbed
gravitational field. Hypothesis H1 tests, under the assumption of
a perturbed gravitational field, whether the ephemeris still provides
a good fit to the observations expected from a perturbation to the
gravitational field. If not, it would contradict the present state
of matters, because the ephemerides do provide good fits to the
data. Hypothesis H2 tests, again under the assumption of a perturbed
gravitational field, whether the model predictions emerging from
the model of a perturbed gravitational field are consistent with
the actual observations. Falsification of either H1 or H2 would
imply that a perturbation to the gravitational field required to
induce the Pioneer anomaly is not present, which under Einstein's
equivalence principle would imply that the Pioneers do not move
geodesically.

The analysis shows that the ephemerides provide good fits to the
perturbed orbits expected from a perturbation to the gravitational
field for Neptune and Pluto, but not for Uranus. Thus hypothesis
H1 is falsified for Uranus. The fact that no significant deviation
in the simulated residuals for Pluto is found can be attributed
to the fact that the measurement uncertainty $\sigma _{l}$ for Pluto
is higher than for the other two planets. Furthermore, the analysis
also shows that the deviation between real and predicted r.a. observations
is statistically significant for both Uranus and Pluto, implying
that hypothesis H2 is falsified for both planets. The statistical
significance of this result is well beyond the 99\% confidence limit
in both cases. For Neptune, on the other hand, we find no significant
deviation, so hypothesis H2 is verified for Neptune. This is not
surprising, because Neptune's orbit has very low eccentricity, and,
as we saw from the discussion of this subject in Section \ref{XRef-Subsection-61013050},
the angular residual arising from the Pioneer effect can be fitted
out in the circular limit (the small perturbation to the observation
angle that comes from projecting the remaining radial residual onto
the celestial sphere is ignored in this study).

It should be stressed again that these conclusions rest on judicial
esitmates of the uncertainties in the observations. We have chosen
to use the standard deviations $\sigma _{l}$ of the real residuals
as the best estimates of the measurement uncertainties.

Based on the available observations of the three outer planets,
an assessment of their uncertainties and the available residuals
between observations and ephemerides, we are therefore lead to the
conclusion that the presence of a perturbation to the gravitational
field necessary to induce the Pioneer anomaly is in conflict with
available data. This implies that the Pioneer anomaly can not have
gravitational origin and, consequently, the Pioneer spacecraft do
not follow geodesic trajectories through space. Hence, any external
interaction model introduced\ \ to explain the Pioneer anomaly should,
in one way or another, violate the equivalence principle. It should
be stressed that ``gravitational origin'' in this context really
means any physical interaction that is mediated indirectly via the
space-time metric only. It is assumed here that particles move geodesically
when moving under gravitational influence only (minimal coupling),
so models implying other, non-minmal coupling schemes can not be
excluded by our results. Our conclusion is consistent with recent
findings of Iorio and Giudice \cite{Iorio-2006a} as well as Izzo
and Rathke \cite{Izzo-Rathke-2005}. It should be noted though, that
while Iorio and Giudice find large secular effects on the orbits
of the three outer planets, we find no such effects. The reason
for this discrepancy is that, as shown here, any secular effects
would be canceled by the simulated ephemeris, i.e. the best fit
solution to the unperturbed equations of motion. The only observable
effect is therefore the net residuals that still remain after the
ephemeris cancels the big secular deviations.

This result can be used to rule out any model of the Pioneer anomaly
that implies that the Pioneer spacecraft move geodesically through
space and that explains the Pioneer anomaly as the effect of a disturbance
to the spacetime metric, regardless of how this disturbance is created.
This includes some of the proposed models of the anomaly mentioned
in section \ref{XRef-Section-215221244}, such as scalar field models\cite{Bertolami-Paramos-2005},
braneworld scenarios \cite{Bertolami-Paramos-2004} and models involving
alternative metric theories of gravitation \cite{Jaekel-Reynaud-2005b,Jaekel-Reynaud-2005,Bekenstein-2004,Bekenstein-Magueijo-2006}
such as TeVeS\cite{Bekenstein-2004}.

We conclude that, in order to resolve the enigma of the Pioneer
anomaly, there is a need for theories and models that may explain
the apparently non-geodesic motion of the Pioneer spacecraft.

\acknowledgments{}

I would like to thank Professor Finn Ravndal at the University of
Oslo for very valuable discussions while preparing this paper. I
am grateful to Dr. E.M. Standish at JPL for providing me with the
residuals of DE 414. I would also like to express my deepest gratitude
for very thorough review reports made by an anonymous referee. The
constructive criticism and numerous valuable suggestions made by
the referee were indeed very helpful in preparing this paper.

\appendix

\section{Pertubative expressions}
\label{XRef-AppendixSection-68175232}

Here, we provide the expansions to third order in eccentricity of
the perturbative expressions given in Section \ref{XRef-Subsection-32891218}
above. All pertubative calculations were made with these third order
expressions, so the results given in this paper are indeed valid
to third order in eccentricity.

The $\mathcal{O}( e^{3}) $ perturbation in reciprocal radial distance,
$\delta u$, which to lowest order is given by\ \ eq. \ref{XRef-Equation-32771920},
is
\begin{multline}
\delta u=\frac{\kappa }{a} \left( 1+\frac{20}{8} e^{2}-\frac{e^{2}}{2}
\cos  2\nu +\frac{e^{3} }{8}\cos  3\nu \right) \\
- \frac{e }{a}\left( 1+\frac{5}{2} e^{2}\right)  \cos  \nu \\
+A \sin ( \sqrt{1+2 \kappa } \nu ) +B \cos ( \sqrt{1+2\kappa }\ \ \nu
) +\mathcal{O}( e^{4})
\end{multline}

Demanding $\delta u=0$ for $\kappa =0$ fixes $A$ and $B$ to take
the values $A=0$ and $B=\frac{e}{a}(1+\frac{5}{2}e^{2})+\mathcal{O}(
e^{4}) $. This gives an angular perturbation
\begin{multline}
\delta u=\frac{\kappa }{a}\left\{  1+\frac{20}{8} e^{2}-\frac{e^{2}}{2}
\cos  2\nu +\frac{e^{3} }{8}\cos  3\nu \right. \\
\left. -e( 1+\frac{5}{2}e^{2})  \nu  \sin  \nu \right\} +\mathcal{O}(
e^{4}) .
\end{multline}

The third order expression for the radial perturbation $\delta r$,
corresponding to eq. \ref{XRef-Equation-44162627}, is
\begin{multline}
\delta r=-\kappa  a\ \ \left\{ 1+2 e^{2}-e\ \ \left( 2+\frac{7}{2}
e^{2}\right)  \cos  \nu + e^{2} \cos  2\nu \right. \\
-\frac{3}{8}e^{3} \cos  3 \nu -e( 1+\frac{5}{4} e^{2} )  \nu  \sin
\nu \\
\left. + e^{2} \nu  \sin  2 \nu -\frac{3}{4}e^{3} \nu \ \ \sin
3 \nu \right\} +\mathcal{O}( e^{4}) .
\end{multline}

Expanding the equation for the angular perturbation to third order
in eccentricity gives
\begin{multline}
\frac{d \delta \varphi }{d\varphi  }= 2\kappa  \left\{ 1+2 e^{2}-
e( 1+2 e^{2})  \cos  \nu  \right. \\
\left. - e \nu ( 1+2e^{2}- e \cos  \nu +\frac{e^{2}}{2}\cos  2\nu
)  \sin  \nu +\frac{1}{8}e^{3} \cos  3\nu \right\} \\
+\mathcal{O}( e^{4}) .%
\label{XRef-Equation-699145}
\end{multline}

Integration of eq. \ref{XRef-Equation-699145}\ \ wrt. $\varphi$
gives the angular perturbation corresponding to eq. \ref{XRef-Equation-44105713}:\ \ \
\begin{multline}
\delta \varphi = \kappa  \left\{ \nu (2 \left( 1+2 e^{2}\right)
+\frac{e}{2}\left( 4+7 e^{2}\right) \ \ \cos  \nu  \right. \\
\left. -\frac{e^{2}}{6}\left( 3 \cos  2\nu -e \cos  3\nu \right)
\right) \\
\left. -\frac{e}{2}\ \ \sin  \nu  \left( 8+15 e^{2} -e \cos  \nu
\right)  +\frac{e^{3}}{36}\sin  3\nu \right\} +\mathcal{O}( e^{4})
.%
\label{XRef-Equation-69151412}
\end{multline}

\section{Terminology and notation}
\label{XRef-AppendixSection-31115539}

Here, we provide a short review of the astonomical terminology and
notation used in the paper. For a complete account of this terminology,
see any astrodynamics textbook, such as \cite{Vallado-2001}.

The {\itshape Ecliptic plane} is the plane in which the Earth moves
around the Sun. The {\itshape Ecliptic} is the cross section of
the Ecliptic plane with the celestial sphere. The {\itshape right
ascension} (r.a.) and {\itshape declination} (dec.) angles fix the
object on the celestial sphere. These angles are measured relative
to the celestial equator, which is the cross section of Earth's
equatorial plane with the celestial sphere. The right ascension,
denoted $\alpha $,\ \ measures the angle of the object eastward
along the celestial equator from a fixed point on the sky; the {\itshape
vernal equinox},\ \ denoted $\Upsilon$, which is the point where
the Sun crosses the celestial equator in the spring. The declination,
denoted $\delta $, measures the angle northward from the celestial
equator. The {\itshape ascending node} is the direction where the
object ascends through the ecliptic. The {\itshape argument of latitude}
(here denoted $\varphi$) is the angular distance of the object measured
in the orbital plane from the ascending node, along the direction
of movement. The {\itshape argument of periapsis} (denoted $\omega$)
is the angular position of the {\itshape periapsis}, which is the
point in the orbit where the object is closest to the Sun. The {\itshape
true anomaly,} denoted $\nu$, is the angular position of the object,
measured from the periapsis, in the orbital plane and in the direction
of movement. $\nu \equiv \varphi -\omega $.

\section{Data used in calculations}
\label{XRef-AppendixSection-4384157}

Table \ref{XRef-TableTitle-4116632}\ \ lists various constants used
in the residual calculations.
\begin{table}
\caption{Data used in the calculations} \label{XRef-TableTitle-4116632}
\begin{ruledtabular}

\begin{tabular}{lccc}
Variable & Uranus & Neptune & Pluto\\
\hline
$t_{0}$ & 1914-07-08 & 1913-12-28 & 1914-01-23\\
  & 06:59:46 & 06:41:17 & 18:58:11\\
$t_{1}$ & 2006-09-30 & 2006-09-30 & 2006-08-26\\
  & 05:45:39 & 04:11:14 & 02:45:13\\
$\nu _{0}$(deg) & -220.57 & 72.49 & $\text{-133.75}$\\
$\nu _{1} (\deg )$ & 171.85 & 273.16 & 40.20\\
$\mathcal{M}_{0} (\deg )$ & -224.19 & 71.53 & -109.47\\
$\mathcal{M}_{1} (\deg )$ & 171.06 & 274.17 & 24.74\\
e & 0.0473 & 0.00882 & 0.251\\
a (AU) & 19.19 & 30.07 & 39.53
\end{tabular}
\end{ruledtabular}
\end{table}

$t_{0}\equiv $ Time of initial data, $t_{1}\equiv $ Time of final
data, $\nu _{0}=\nu ( t_{0}) =$ true anomaly at time of initial
data (deg), $\nu _{1}=\nu ( t_{1}) =$ true anomaly at time of final
data (deg), $\mathcal{M}_{0}=\mathcal{M}( t_{0}) =$ mean anomaly
at time of initial data (deg), $\mathcal{M}_{1}=\mathcal{M}( t_{1})
=$ mean anomaly at time of final data (deg).\label{Def-gravitational}\label{constant-kappa}\label{Pluto-data-remark}\label{best-fit}\label{WhyNotGaussEquations}\label{why-argumentoflatitude}

\end{document}